%% file: RPDNN_lrec2020.tex
\documentclass[10pt, a4paper]{article}
\usepackage{lrec}
\usepackage{multibib}
\newcites{languageresource}{Language Resources}
\usepackage{graphicx}
\usepackage{tabularx}
\usepackage{soul}

\usepackage{epstopdf}
\usepackage[latin1]{inputenc}

\usepackage{hyperref}
\usepackage{xstring}
\usepackage{color}
\usepackage{todonotes}

\usepackage{url}
\usepackage{lipsum}
\usepackage[]{multirow}
\usepackage{enumitem}
\usepackage{float}
\usepackage{caption}
\usepackage{subcaption}

\title{RP-DNN: A Tweet level propagation context based deep neural networks for early rumor detection in Social Media}

\name{Jie Gao, Sooji Han, Xingyi Song, Fabio Ciravegna}

\address{Regent Court, 211 Portobello, Sheffield, UK, S1 4DP \\
         \{j.gao, sooji.han, x.song, f.ciravegna\}@sheffield.ac.uk\\}

\abstract{Early rumor detection (ERD) on social media platform is very challenging when limited, incomplete and noisy information is available. Most of the existing methods have largely worked on event-level detection that requires the collection of posts relevant to a specific event and relied only on user-generated content. They are not suitable for detecting rumor sources in the very early stages, before an event unfolds and becomes widespread. In this paper, we address the task of ERD at the message level. We present a novel hybrid neural network architecture, which combines a task-specific character-based bidirectional language model and stacked Long Short-Term Memory (LSTM) networks to represent textual contents and social-temporal contexts of input source tweets, for modelling propagation patterns of rumors in the early stages of their development. We apply multi-layered attention models to jointly learn attentive context embeddings over multiple context inputs. Our experiments employ a stringent leave-one-out cross-validation (LOO-CV) evaluation set-up on seven publicly available real-life rumor event data sets. Our models achieve state-of-the-art(SoA) performance for detecting unseen rumors on large augmented data which covers more than 12 events and 2,967 rumors. An ablation study is conducted to understand the relative contribution of each component of our proposed model.
 \\ \newline  }

\begin{document}

\maketitleabstract


\input{sections/introduction}
\input{sections/relatedwork}
\input{sections/method}
\input{sections/experiments.tex}
\input{sections/results}
\input{sections/conclusions.tex}

\section{Bibliographical References}
\label{main:ref}
\bibliographystyle{lrec}
\bibliography{RPDNNlrec2020}

\section{Language Resource References}
\label{lr:ref}
\bibliographystylelanguageresource{lrec}
\bibliographylanguageresource{RPDNNlrec2020}

\input{sections/appendix_model_setting.tex}
\input{sections/appendix_loocv_details.tex}
\input{sections/appendix_attention_analysis.tex}

\end{document}

%% file: sections/introduction.tex
\section{Introduction}\label{intro}

Research on social media rumors has become increasingly popular to understand the emergence and development of rumor events. An automatic and efficient approach for the early identification of rumors is vitally necessary in order to limit their spreading and minimize their effects. 

A typical rumor resolution process can include four sub-tasks: rumor detection, tracking, stance classification, and verification \cite{zubiaga18detection}. Rumor detection which aims to identify whether a claim is a rumor or non-rumor is a fundamental task for rumor resolution. Once a rumor is identified, it becomes possible to track its evolution over time, identify its sources, perform stance detection, and finally determine the its veracity \cite{zubiaga18detection} \citelanguageresource{kochkina2018all}. Recent research on online rumors has largely focused on the later stages of the process, that is, stance classification and verification. Although these are crucial for rumor resolution, they cannot be performed until rumors are identified. Several studies skip this preliminary task, either leaving the development of approaches for them for future work or assuming that rumors and their associated posts are manual inputs. In this work, we highlight the importance of developing automated ERD systems for the success of the entire rumor resolution process.

We propose a hybrid and context-aware deep neural network framework for \textit{tweet-level} ERD, which is capable of learning not only textual contents of rumors, but more importantly social-temporal contexts of their diffusion. A large body of SoA research on rumor detection \cite{lukasik2015classifying,chen2018call,zhou2019early} only leverages language modeling at the word level for contents of source tweets and contexts (typically replies). In contrast, we pay more attention to modeling at social context level. Social contextual information typically refers to conversational threads of source tweets such as replies and retweets in the case of Twitter. Conversational threads provide time series information that how rumor-mongering changes people's opinions and how social media allows self-correction. Some research uncovers two competing rules including majority preference and minority avoidance that affect the evolution of public opinion through information exchange \cite{wang2017rumor}. Therefore, conversational threads offer valuable insights about rumor propagation at the single tweet level before events become widespread and obtain far-reaching impact. 

Twitter metadata provides rich explicit and implicit cues related to replies and retweets (e.g.,author information, decay of interest, and chain of replies) which can provide useful complementary signals for early diffusion and have the potential advantage of platform, domain and language portability. Different from most existing work which is exclusively based on textual contents, we argue that a good model for temporal sequence learning can benefit from multiple inputs. Multi-modal temporal data can offer different representations of the same phenomenon. In the case of content and metadata in conversational threads, they are correlated and share high-level semantics \cite{kiciman2010language}. Motivated by this observation, our method aims to extend a model based on rumor source content (SC) with social context information. A SoA context-aware Neural Language Model (NLM) fine-tuned specifically for the task of rumor detection is employed to encode contents. Social contexts are modeled as the joint representation of conversational contents and metadata through a Recurrent Neural Network (RNN) architecture. We leverage two types of complementary contextual information which are strongly correlated with source tweet contents. Specifically, we utilize social context content (CC) to provide insights about how public opinion evolves in early stages and social context metadata (CM) to provide auxiliary information on how rumors spread and how people react to rumors.


The main contributions of this work can be summarized as follows: 

(1) We propose a hybrid deep learning architecture for rumor detection at the \textit{individual tweet level}, while the majority of recent work focuses on \textit{event-level} classification. It advances SoA performance on \textit{tweet-level} ERD. 

(2) We exploit a \textit{context-aware} model that learns a unified and noise-resilient rumor representation from multiple correlated context inputs including SC, CC and CM beyond the word-level modeling via a rumor task-specific neural language model and multi-layered temporal attention mechanisms.  

(3) A large, augmented rumor data set recently released \citelanguageresource{han2019neural} is employed to train our proposed model. Extensive experiments based on an ablation study and \textit{LOO-CV} are conducted to examine its effectiveness and generalizability. Our model outperforms SoA models in \textit{tweet-level} rumor detection and achieves comparable performance with SoA \textit{event-level} rumor detection models.
\textbf{}

%% file: sections/relatedwork.tex
\section{Related Work}\label{relatedwork}

There are two different objectives in most recent techniques proposed to date, including 1) {\bf event-level rumor detection}: its purpose is to classify the target event into rumor and non-rumor. It involves story or event detection and tracking as well as grouping retweets or similar tweets in clusters during pre-processing \cite{chen2018call,kwon17rumor,ma2016detecting,guo2018rumor,nguyen17early,Jin17Rumor,wang2018eann}. 2) {\bf tweet-level detection}: in contrast to the event-level detection, it aims to detect individual rumor-bearing source tweets before events unfold \cite{zubiaga2016analysing}. This paper focuses on tweet-level detection. This is more challenging work than the event-level detection because individual tweets are short, noisy, and of divergent topics due to intrinsic properties of social media data. Thus, modeling tweet-level ERD with limited context is still considered as open issue \cite{zubiaga18detection}.

\textbf{Event-level rumor detection}  \cite{yu17convolutional} proposes a CNN-based misinformation detection architecture which allows CNNs to learn representations of contents of input tweets related to an event. \cite{ma2016detecting} proposes various models based RNNs which learn tweet content representations based on tf-idf. \cite{ruchansky17csi} proposes a framework which jointly learns temporal representations and user features of input posts. \cite{ma2018detect} proposes a GRU-based, multi-task learning architecture which unifies both stances and rumor detection. \cite{chen2018call} is one of early work that uses RNNs and attention mechanism to model deep representation of aggregated tweet content of rumor event. \cite{guo2018rumor} exploits content representations and hand-crafted social contexts features with attention-based bidirectional LSTM networks. 

\textbf{Message-level rumor detection} \cite{zubiaga17exploiting} proposes a conditional random fields-based model that exploits a combination of context content and metadata features to learn sequential dynamics of rumor diffusion at the tweet level. \cite{ma18rumor} proposes recursive neural networks models which take a tree structure of each input source tweet as input. Tree structures represent relations between source tweets and their contexts (i.e., replies and retweets). \cite{liu2018early} proposes a hybrid of CNNs and RNNs which is capable of learning rumor propagation based on features of users who have participated in rumor spreading. \cite{jin17multimodal} proposes a multi-modal model comprising CNN and LSTM with attention mechanism. It jointly learns representations of rumour textual contents and social contexts. The joint representations are fused with images embedded in tweets encoded using CNNs.  A recent trend is to exploit multi-task learning frameworks for rumor detection and other rumor resolution sub-tasks  \citelanguageresource{kochkina2018all} \cite{veyseh19rumor,li19rumor}. The majority of such work focuses on leveraging tweet content representation and the conversational structure of their context (e.g., replies). \citelanguageresource{kochkina2018all} decomposes conversation threads into several branches according to Twitter mentions (i.e., @username) which allows the application of majority voting for per-thread prediction. \cite{veyseh19rumor} examines the effectiveness of recent NLMs in content embedding. \cite{li19rumor} incorporates user-level information as an additional signal of credibility. \cite{geng19rumor} incorporate the sentiment of replies into their GRU model and applies self-attention to source tweet content. \cite{han2019data} modified the RNN-based multi-task learning model originally proposed by \citelanguageresource{kochkina2018all}. The authors evaluate the proposed model using their augmented data generated via weak supervision.

In this paper, we identify several limitations of existing work on tweet-level rumor detection. The majority of SoA methods are limited to contents of source posts and/or those of their contexts and rely on hand-crafted features for both content and propagation context. Our work avoids any sophisticated feature engineering on content and only adopts a limited number of generic features commonly used to encode context metadata. In addition, prevalent word-level attention mechanism is not applied in our model. This helps us to focus on examining the effectiveness of our propagation context-based model and task-specific language model. Furthermore, data scarcity is a known limitation in the field of ERD. Most studies have evaluated their methods on small or proprietary data sets with a conventional approach for splitting data into train and test sets. To our best knowledge, it is the first work presenting an extensive experimental comparison with both LOO-CV and k-fold CV procedures to provide an almost unbiased estimate of the generalizability of a model to unseen events and in realistic scenarios.

%% file: sections/method.tex
\section{Methodology}\label{method}
\subsection{Problem Statement}\label{method:problemstatement}
Rumors are commonly considered as statements presenting facts that lack substantiation. Therefore, candidate rumor tweets should be factual or informative. In our task, a potential rumor is presented as a tweet which reports an update associated with a newsworthy event, but is deemed unsubstantiated at the time of release. Individual social media posts can be very short in nature, containing very limited context with variable time series lengths. This is a typical characteristic on Twitter. A rumor claim in the very early stages of event evolution is usually from a candidate source tweet $x_i$ at timestamp $t_i$, which can be considered as a source of a potential rumor event. In this paper, we focus on conversational content and associated metadata which are considered as two separate but correlated sequential sub-events.


A set of candidate source tweets is denoted by $X=\{x_1, ..., x_n\}$ which contains $i$ candidate tweets, where each candidate tweet $x_i = \{[CC_{i}, CM_{i}], t_{i}\}, x_i \in X$ consists of two correlated observations (reactions) $CC_{i}$ and $CM_{i}$ over time series $t_{i}$. Let $j$ be the length of conversational threads (i.e., the number of replies) of each input source tweet. $CC_{i} = \{cc_{i,0}, cc_{i,1}, ... ,cc_{i,j} \}$ is a set of temporal-ordered observations from {\bf context content}. $CM_{i} = \{cm_{i,0}, cm_{i,1}, ..., cm_{i,j}\}$ is a set of temporal-ordered observations from {\bf context metadata}. Let $y = \{0, 1\}$ be binary labels. The task is to predict the most probable tag for each candidate source tweet $x_i$ based on source tweet content and all context sub-events $CC_{i}$ and $CM_{i}$, given a time range $t_i \subseteq [0, j]$. $y_i=1$ if $x_i$ is a rumor, and $y_i=0$ otherwise.

\subsection{Overview of Model Architecture}\label{modelArchit}
\begin{figure*}[htb!]
\centering
\includegraphics[scale=0.45]{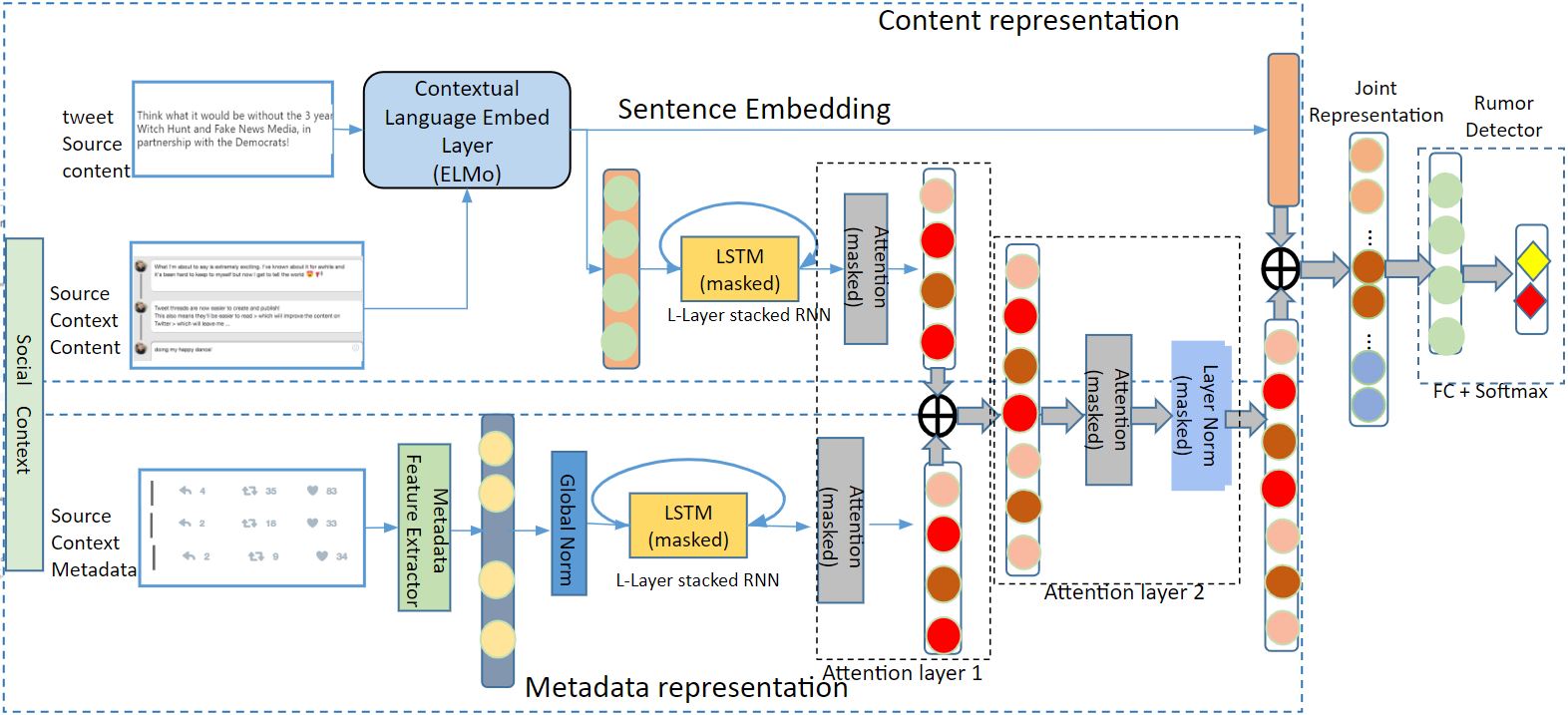}
\caption{Overview of model architecture}\label{fig:overview}	
\end{figure*}

The overall architecture of the proposed tweet-level \textit{Rumor Propagation based Deep Neural Network} (RP-DNN) is shown in Figure~\ref{fig:overview}. Basically, we learn a neural network model that takes source tweets $x_i$ and corresponding contexts ($CC_{i}$ and $CM_{i}$) as input and outputs predictions $\hat{y_i}$. RPDNN consists of four major parts including 1) data encoding layers, 2) stacked RNN layers, 3) stacked attention models, and 4) classification layer. 

Tweet-level EDR using RP-DNN follows the four key stages: \textbf{a)} Once candidate source tweets $X$ and associated context inputs ($CC_{i}$ and $CM_{i}$) are loaded and pre-processed (see details in section \ref{exp:datapreprocessing}), the two types of raw context inputs will be encoded in \textbf{\textit{data encoding layers}}. These are important layers that convert source tweets and conversational context into inputs for subsequent RNN layers for contextual modeling. It consists of a content embedding layer (section \ref{method:cc_encoder}) and a metadata encoding layer (section \ref{method:contextmetadata}). The objective of the former is to convert tweets into embeddings $V_{cc}^i$. The latter is to use a Metadata Feature Extractor (MFE) to extract features from the corresponding metadata of the tweets that characterizes public engagement and diffusion patterns. The output of the MFE is represented as feature vectors $V_{cm}^i$ which are normalized by applying a global mean and variance computed from training data. \textbf{b)} Subsequently, encoded context inputs will be fed into a social-temporal context representation layer consisting of \textbf{\textit{stacked RNN layers}} and \textbf{\textit{stacked attention models}} (illustrated in section \ref{method:contextencoder} and \ref{method:stackedSoftAttention} respectively). We stack multiple LSTMs together to form a stacked LSTM that takes input representations (i.e., $V_{cc}^i$ and $V_{cm}^i$; outputs of the data encoding layers) arranged in chronological order. Let the number of layers be $L$. L-layer LSTMs ($L=2$ in our case) are utilized to process the two types of contextual data separately. \textbf{c)} The recurrent structure models features of sequential data and then uses soft hierarchical attention models (the 1st attention layer) to produce an optimal representation. The contextual embeddings from the two recurrent layers (hidden states) output ($H_{cm}^i$ and $H_{cc}^i$) are then temporally combined to form a joint representation ($H_{c}^i$). The third attention model (the 2nd attention layer) is performed on the joint hidden sequential embedding $H_{c}^i$ and eventually produces a compact representation of context sequences $V_c^i$, followed by (masked) layer normalisation \cite{ba2016layer}. \textbf{d)} Finally, we combine two embeddings of SC and context via concatenation to form the final rumor source representation in the \textbf{\textit{classification layer}}. This is the final output layer which provides the result of rumor detection. Cross-entropy loss are computed to optimize the whole network. A 3-layer fully-connected neural network with Leaky ReLu activations and softmax function takes the final representation to yield the output.

\input{sections/stacked_rnns.tex}
\subsection{Stacked Soft Attentions}\label{method:stackedSoftAttention}

In order to amplify the contribution of important context elements and filter noise or unnecessary information in final representation, we introduce multiple-layer stack attention mechanism in our network. This is inspired by the performance of stacked attentions in recent advances \cite{dyer2015transition,yang2016stacked}. By applying attention over multiple steps, the model can focus on more salient features and this has been proved in many visual recognition challenges \cite{yang2016stacked}. We explore ways to leverage attention mechanisms for context embeddings at different levels to eliminate invalid information and get more accurate contextual interaction information, thereby improving classification performance.

Specifically, we propose to calculate attention weights by providing  information about all time steps for context embedding layers. It takes a context sequence of a predefined length $j$ as input and learns a mapping from this sequence to an output sequence using attention mechanisms. We employ the idea of hierarchical attention networks \cite{yang2016hierarchical} and adapt the context-aware model in our networks. We here represent attention as a probabilistic distribution over temporally ordered conversational context inputs, and implement its estimation via our end-to-end rumor classification framework. The standard softmax function \cite{martins2016softmax} is used to approximate a normalized probability distribution of importance on entire context. Let $H_c$ be the recurrent hidden states of tweet context (see section \ref{method:contextencoder}). Formally,

\begin{equation} \label{eq:attention1} 
\alpha_c^t = softmax(tanh(W_hh_c^t + b_h)), \forall t \in [0, j].
\end{equation}

\begin{equation} \label{eq:attention2}
h_{c\_new}^t = \alpha_c^th_c^t
\end{equation}

$W_h$ and $b_h$ are the attention layer's weights, which are initialized using He initialization and optimized during training. Zero padding is used to handle variable lengths. Following the same practice adopted in the stacked RNN layer, we mask out padded values with negative infinity float following the practice of \cite{vaswani2017attention}. $h_{c\_new}$ is the re-weighted context embeddings.

Rather than only computing attention weights once, attention mechanism is applied to two layers in our architecture: 1) stacked RNN layers and 2) joint representation layer. Specifically, the first attention layer contains two sub-layers of attentions on the top of CC context encoder (see section \ref{method:cc_encoder}) output $H_{cc}^t$ and CM context encoder (see section \ref{method:contextmetadata}) output $H_{cm}^t$ respectively (as defined in eq. \ref{eq:cc_attention} and \ref{eq:cm_attention}). Two independent attention models are trained and then modify the hidden states of two separate recurrent layers. The output of two attention models are denoted as $H_{cc\_new}^t$ and $H_{cm\_new}^t$. The weighted hidden state vectors for all time-steps from two context encoders are then concatenated and provided as joint representation input for the second attention layer. 

\begin{equation} \label{eq:cc_attention}
H_{cc\_new}^t = attention_1(H_{cc}^t)
\end{equation}

\begin{equation} \label{eq:cm_attention}
H_{cm\_new}^t = attention_1(H_{cm}^t)
\end{equation}

To determine the inference relationship between two correlated context embeddings and to verify our hypothesis, we use the attention model as a composition layer to mix the two types of sub-event inference information. Different from the first attention layer, the second attention layer aggregates all the hidden states using their relative importance via weighted sum, which is trained in the hope to capture shared semantics between content and metadata. Eventually, the proposed algorithm helps to incorporate additional auxiliary information into a unified representation of reaction and diffusion patterns to achieve outstanding performance in our context-based EDR problem. Formally,

\begin{equation} \label{eq:joint_attention}
h_{c}^t = attention_2(h_{cc\_new}^t \oplus h_{cm\_new}^t)
\end{equation}

\begin{equation} \label{eq:weighted_sum_att}
v_{c} = \sum_{t}h_{c}^t
\end{equation}

where $h_{c}^t$ is the joint hidden states of context and $v_{c}$ is the final context vector, i.e., a sum of $h_{c}^t$ for all time steps.




\subsection{Tweet content encoder}\label{method:cc_encoder}
A large body of work \cite{zubiaga18detection} has previously proposed and demonstrated the effectiveness and advantage of textual contents in rumor detection. The user-generated content has been proved to be useful for providing effective signals for identifying emerging rumors. For instance, credibility-related terms (e.g., "reportedly", "I hear that", etc.) can effectively indicate the uncertainty of a candidate tweet \cite{zubiaga17exploiting}. For rumor tweets that do not have sufficient signals, social context content can provide information about how people react to them, which has been exploited extensively to identify rumors and their veracity. \cite{maddock2015characterizing,zubiaga18detection}. 

In our framework, tweet content embeddings are obtained via ELMo \cite{Peters2018deep}, a SoA context-aware neural language model (NLM). We allow a NLM to learn signals for linguistic and semantic characteristics from rumor tweet content such as ambiguity and uncertainty in order to avoid using hand-crafted features. ELMo represents each individual word while taking the context of an entire corpus (e.g., a sentence and paragraph) into account. The weight of each hidden state is task-specific and can be learned from domain-specific corpora. In our architecture, tweet sentence embeddings are learned from both domain-specific and general corpora. We employ a SoA ELMo model fine-tuned specifically for the task of rumor detection on social media \citelanguageresource{han2019neural}. This domain-specific language model is pre-trained on an 1 billion word benchmark corpus with vocabulary of 793,471 tokens and then fine-tuned on a large credibility-focus Twitter corpus with 6,157,180 tweets with 146,340,647 tokens and 2,235,075 vocabularies. The fine-tuned model achieves low perplexity in in-domain data sets and SoA performance in  the rumor detection task. Following the practice in \cite{perone2018evaluation,han2019data}, averaging ELMo word vectors is employed to produce the final short-text embeddings, using features from all three layers of the ELMo model.

\input{sections/method_cxt_features.tex}

%% file: sections/stacked_rnns.tex
\subsection{Stacked RNN layer}\label{method:contextencoder}
A natural choice is to use Recurrent Neural Network (RNN) to model rumor context. An RNN processes a sequential input in a way that resembles how humans do it. It performs an operation, $h_t = f_W (x_t, h_{t-1})$, on every candidate tweet context ($x_t$) of a sequence, where $h_t$ is the hidden state a time step $t$ and $W$ is the weights of the network. The hidden state at each time step depends on the previous hidden state. Therefore, the order of time series-based reaction context input is important. Intuitively, this process enables RNNs to model the evolution of public opinion about each source claim and diffusion patterns of public engagement (e.g., retweets, likes) through corresponding metadata. Meanwhile, it enables to handle inputs of variable lengths.

Regarding utilizing complementary context clues and modeling context with different types of features (considered as two different sub-events), conventional approaches \cite{xing2017incorporating,zhou2017disambiguating,jin17multimodal,gu2018deep} simply concatenate embeddings of different data inputs or process them through a linear combination of different feature embeddings to form a single representation. This practice completely ignores the correlations and differences between different context inputs. We argue that a model should have the ability of learning weights separately from different context inputs in order to find salient parts of each context type. In addition, a model should have the ability to learn important clues across multiple context observations (as illustrated in section \ref{method:stackedSoftAttention}). 

To this end, we propose two (simultaneous) context embeddings to explore two correlated context inputs, and use two layers of forward LSTMs in order to learn more abstract features respectively. Concretely, to model the temporal evolution of public opinions, context content embeddings ($V_{cc}^i$) are given as input to two layers of forward LSTMs. The context output state $H_{cc}^i$ at time $t$ is abbreviated as $ \overrightarrow{h_{cc,t}^i} = \overrightarrow{LSTM_{l}}(\overrightarrow{h_{cc,t-1}^{i}},v_{cc,t}^i), \forall t \in [0, j].
 $


Regarding diffusion patterns of public engagement, we employ shallow features extracted from explicit information in social reactions to induce a hierarchical RNN model. In contrast to previous work \cite{ma2015detect,zubiaga17exploiting}, our RNN-based method avoids painstakingly complicated feature engineering, and instead allows RNN to learn deep, hidden behavioural, and social dynamics of \textbf{}underlying complex hierarchical social-temporal structure. The context output state $H_{cm}^i$ at time $t$ is abbreviated as $ \overrightarrow{h_{cm,t}^i} = \overrightarrow{LSTM_{l}}(\overrightarrow{h_{cm,t-1}^{i}},v_{cm,t}^i), \forall t \in [0, j].
 $


%% file: sections/method_cxt_features.tex
\subsection{Conversational Context Metadata}\label{method:contextmetadata}
The proposed architecture leverages 27 hand-crafted and generic features (described in Table \ref{tab:features}) that can be categorized into tweet-level and user-level. 
Early work on rumor detection employs supervised learning techniques, and thus has extensively studied manually curated features related to contents, users, and networks to seek distinguishing features of online rumors \cite{qazvinian11rumour,kwon17rumor,yang12automatic,sun13detecting,zhao15enquiring,zhang2015predictors,wu2015false,ma2015detect,liu16detecting,zubiaga17exploiting,hamidian16rumor}. These studies have shown that those features have the potential for distinguishing rumors from non-rumors. In recent advances of deep learning architectures, few event-level detection techniques \cite{ruchansky17csi,kwon17rumor,liu2018early,guo2018rumor} have shown the merits of combining both hand-crafted metadata features and deep-learned features.



\begin{table}[t!]
\caption{Description of hand-crafted features.}
\label{tab:features}
\scriptsize
\centering
\begin{tabular}{
				p{\dimexpr 0.95\linewidth-1.2\tabcolsep}
				}
\hline
\multicolumn{1}{c}{\textbf{Tweet-level features}} \\ \hline
Number of retweets \\
Number of favorites \\
Whether tweet has a question mark \\
Whether tweet is a duplicate of its source\\
Whether tweet contains URLs\\
Number of URLs embedded in tweet \\
Whether tweet has native media\emph{*}\\
Number of words in tweet except source author's screen name\\
\hline
\multicolumn{1}{c}{\textbf{User-level features}} \\ \hline
Number of posts user has posted \\
Number of public lists user belongs to\\
Number of followers\\
Number of followings\\
Whether user has a background profile image \\
User reputation (i.e., followers/(followings+1))\\
User reputation (i.e., followers/(followings+followers+1))\\
Number of tweets user has liked so far (aka "user favorites")\\
Account age in days\\
Whether user is verified\\
User engagement (i.e., \# posts / (account age+1))\\
Following rate (i.e., followings / (account age+1))\\
Favorite rate (i.e., user favorites / (account age+1))\\
Whether geolocation is enabled\\
Whether user has a description \\
Number of words in user description\\
Number of characters in user's name including white space\\
Whether user is source tweet's author\\
Response time decay (time difference between context and source tweet in mins)\\
\hline
\emph{*} multimedia shared with the Tweet user-interface not via an external link 
\end{tabular}
\end{table}

\noindent$\bullet$ \textbf{Tweet-level features}
We let unsupervised NLM automatically learn syntactic and semantic representations of input tweets. Therefore, our hand-crafted features related to content mainly include features related to URLs and multimedia embedded in tweets. Twitter users often use URLs as additional references due to a length limit \cite{qazvinian2011rumor}. Including them in tweets tends to encourage more people to share rumors \cite{tanaka2014impact} and increase the trustworthiness of tweets  \cite{gupta2012credibility,castillo2011information}. In particular, \cite{friggeri2014rumor} reports that unverified information with links to websites for validating and debunking rumors often goes viral on social media. 

\noindent$\bullet$ \textbf{User-level features}\label{cxt_features_user}
Rumor spreaders are individuals who seek attention and reputation \cite{sunstein2010rumours}. Features related to user profiles and reactions contribute to the characterization of rumors \cite{liu2015real}. Previous studies found that rumors tend to spread from low-impact users to influencers, whereas non-rumors have the opposite tendency \cite{ma2017detect,kwon17rumor}. Another study reports that trustworthy sources such as mainstream media and verified users participate in rumor spreading by simply sharing rumors and maintaining neutrality \cite{li16user}. 

%% file: sections/experiments.tex
\section{Experiments}\label{experiments}
In this section, we report the evaluation data set and methods for our proposed model and data processing methods. 


\subsection{Data sets}
Table \ref{tab:dataset_statistics} presents the statistics of all the pre-filtered event data sets used in our experiment. They are obtained from three public data sets. ``Avg. tdiff'' stands for the average time length of context (conversational threads) in each event data set in minutes.


\textbf{1. Aug-rnr~\citelanguageresource{han2019neural}}: This is an augmented version of the \textit{PHEME (6392078)}. It contains rumor and non-rumor source tweets and their contexts associated with six real-world breaking news events. Source tweets are labeled with weak supervision. The augmented data set expands original one by 200\% of source tweets and 100\% of social context data. The temporal filtered version 2.0 data\footnote{\url{https://zenodo.org/record/3269768}} is adopted in our experiments to examine our models' performance in the context-based ERD task. We only use replies (i.e., context data) posted within 7 days after corresponding source tweets were posted. Retweets are excluded \footnote{Our preliminary results shows that retweets metadata is very noisy. Simply adding retweets into context causes underfitting and poor performance.}.\\
\textbf{2. Twitter15/16 \cite{ma2017detect}}: These two data sets consist of rumor and non-rumor source tweets and their context. The context of each source tweet is provided in the form of propagation trees. Source tweets are manually annotated with one of the following four categories: non-rumor, false rumor, true rumor and unverified rumor. As we restrict the experiment set-up to binary classification, all but ``non-rumor'' class are aggregated into ``rumor'' class. We collect context data by following the practice introduced in \cite{han2019data}. \\
\textbf{3. PHEME (6392078; \citelanguageresource{kochkina2018all})}: This consists of manually labeled rumor and non-rumor source tweets and their replies for 9 breaking news events. It is used to generate test sets during evaluation.

\begin{table}
	\caption{Statistics of 12 events data sets.}
	\label{tab:dataset_statistics}
	\scriptsize
	\centering
	\begin{tabular}{p{\dimexpr 0.15\linewidth-1.2\tabcolsep}|
			p{\dimexpr 0.11\linewidth-1.2\tabcolsep}
			p{\dimexpr 0.125\linewidth-1.2\tabcolsep}|
			p{\dimexpr 0.12\linewidth-1.2\tabcolsep}
			p{\dimexpr 0.06\linewidth-1.2\tabcolsep}
			p{\dimexpr 0.05\linewidth-1.2\tabcolsep} 
			p{\dimexpr 0.06\linewidth-1.2\tabcolsep} 
			p{\dimexpr 0.06\linewidth-1.2\tabcolsep} 
			p{\dimexpr 0.1\linewidth-1.2\tabcolsep} }
		
		\textbf{Event} &   &  & \multicolumn{6}{c}{\textbf{Replies}} \\ 
		\hline
		&  \textbf{\# of\newline rumors} & \textbf{\# of\newline non-rumors}  & \textbf{Total} & \textbf{Avg.} &\textbf{Min} & \textbf{Max} & \textbf{Mdn} & \textbf{Avg. tdiff} \\ \hline 
		charlie & 382 & 1,356 & 42,081 & 24 & 6 & 341 & 19 & 8.6\\
		ferguson & 266 & 746 & 26,565 & 26 & 6 &  288 & 18 & 47.3\\   
		german & 132 & 122 & 4,163 & 16 & 6 & 109 & 14 & 12.8 \\ 
		sydney & 480 & 784 & 26,435 & 21 & 6 & 341 & 17 & 7.1\\
		ottawa & 361 & 539 & 16,034 & 18 & 6 & 208 & 13 & 440.6\\ 
		boston & 75 & 584 & 23,210 &35 & 6 & 207 & 20 & 8.1\\
		ebola & 13 & 0 & 208 & 16 & 6 & 26 & 15 & 42.6\\
		gurlitt & 1 & 1 & 23 & 12 & 7 & 16 & 12 & 174.1\\
		prince & 43 & 0 & 452 & 11 & 6 & 21 & 10 & 4.7\\
		putin & 22 & 9 & 379 & 12 & 6 & 25 & 10 & 21.1\\
		twitter15 & 782 & 323 & 47,324 & 43 & 6 & 458 & 28 & 2.2\\
		twitter16 & 410& 191 & 27,732 & 46 & 6 & 458 &29 & 16.6\\ \hline
		\textbf{Total} & 2,967 & 4,655 & 214,606 & & & & &\\
		
	\end{tabular}
\end{table}

\subsection{Data Preprocessing}\label{exp:datapreprocessing}
In this task, a candidate source tweet has to satisfy the following constraints: (1) \textit{informativeness}: the length of its content (i.e., the number of tokens) should be greater than a minimum value. Tweets that lack enough textual information are generally unremarkable and add noise to data~\cite{Ifrim2014Event}. (2) \textit{popularity}: its context size (i.e., the number of replies to it) should be greater than a minimum value. This pre-filtering allows us to examine the focus of this paper regarding conversational context.  Therefore, each input $x_i$ (i.e., a candidate tweet) is set to satisfy both minimum content length ($=4$) and minimum context length($=5$). All tweets are lowercased, tokenized and deaccented.

\subsection{Model Implementations}
Models were implemented\footnote{The source code is available at \url{https://github.com/jerrygaoLondon/RPDNN}} using Python 3.6, Allennlp (0.8.2) framework\cite{gardner2018allennlp}, and Pytorch 1.2.0. All models were trained on one Tesla P100 SXM2 GPU node with maximum 16GiB RAM. More details of model settings are given in appendix \ref{appendix_model_settings}.

\subsection{Settings and Baselines}
Two following evaluation procedures are employed to evaluate our models. Four performance metrics are adopted in our experiments including Accuracy (Acc.), precision (P), recall (R), and F1-measure ($F_1$). P, R and $F_1$ are computed with respect to positive class (i.e., rumor). Overall performance is an average over all CV folds.

\textbf{LOO-CV} The mainstream rumor detection methods \cite{ma2016detecting,liu2018early,chen2018call,ma18rumor,zhou2019early,tarnpradab2019attention} adopt conventional K-fold Cross Validation (CV) procedures with various different split ratios to estimate their models' performance. This practice allows similar distributions between train and test sets, and usually leads to good performance. However, the simple train/test split seems weak when a model is required to generalize beyond the distribution sampled from the same rumor event data. To this end, we adopt \textit{Leave one (event) out cross validation (LOO-CV)} as an approximate evaluation of our proposed models in realistic scenarios.

Our LOO-CV data is presented in Table \ref{tab:loocv_statistics}. 12 real-world rumor event data sets in total are used to generate balanced training, hold-out and test data. Two types of samples (i.e., rumor and non-rumor) are randomly shuffled in each data set. Training and hold-out sets contain augmented data sets from \textit{Aug-rnr}, generated from 11 (out of 12) events with a split ratio 9:1. 7 manually labeled event data sets from \textit{PHEME (6392078)} and \textit{Twitter15/16} are selected as test sets, thus it is 7-fold LOO-CV.

\begin{table}
\caption{Statistics of the balanced data sets for LOO-CV.}
\label{tab:loocv_statistics}
\footnotesize
\centering
\begin{tabular}{p{\dimexpr 0.25\linewidth-1.2\tabcolsep}
				p{\dimexpr 0.16\linewidth-1.2\tabcolsep}
				p{\dimexpr 0.18\linewidth-1.2\tabcolsep}
				p{\dimexpr 0.11\linewidth-1.2\tabcolsep}
			     }
LOO Event &  \textbf{Training} & \textbf{Hold\-out}  & \textbf{Test}  \\ \hline 
charlie  & 4,674 & 496 & 680 \\
ferguson & 4,818 & 584 & 466 \\   
german & 5,144 & 526 & 212 \\ 
sydney & 4,474 & 500 & 836 \\
ottawa & 4,676 & 536 & 578 \\ 
twitter15 & 3,924 & 446 & 646 \\
twitter16 & 4,600 & 514 & 382 \\

\end{tabular}
\end{table}

\textbf{K-fold CV} We also evaluate our models via 5-fold cross validation following the common practice in this field in order to provide a comparative evaluation with more SoA methods. Stratified k-fold CV is employed to ensure that the percentage of samples for each class is preserved in each returned stratified fold. The split ratio for three data sets is 18:1:1, which results in 4,382 source tweets in the training set and 246 in hold/test set per fold. 


\textbf{Baselines} Our models (see Section \ref{exp:ablation_study}) are evaluated with the following SoA models that are comparable and utilize conversational threads.

\noindent$\bullet$ \textbf{\cite{zubiaga17exploiting}: } LOO-CV results for tweet-level classification on positive class (i.e., rumor) are given on 5 PHEME event sets. 

\noindent$\bullet$ \textbf{\cite{zhou2019early}: } Overall results of event-level ERD for two classes with a 3:1 train/test split ratio are provided for the 5 PHEME event sets .  

\noindent$\bullet$ \textbf{\cite{han2019data}: } LOO-CV results for tweet-level ERD on the 5 PHEME event sets are provided based on a train/test split ratio of 3:1.

\noindent$\bullet$ \textbf{\cite{ma18rumor}: } 5-fold CV results for tweet-level ERD for four classes on the Twitter 15/16 are available with a 3:1 train/test split ratio.

\noindent$\bullet$ \textbf{\cite{liu2018early}: } 3-fold CV results for event-level ERD for two classes are reported on the Twitter 15/16 with a 3:1 train/test split ratio.

\subsection{Ablation study}\label{exp:ablation_study}

A set of exploratory experiments is conducted to study the relative contribution of each component in our message-level ERD model. 

\noindent$\bullet$ \textbf{RPDNN}: This is our full model setting that we will compare with baseline methods.

\noindent$\bullet$ \textbf{RPDNN-cxt}: Only source contents are used. 

\noindent$\bullet$ \textbf{RPDNN-SC}: Only social contexts are used.

\noindent$\bullet$ \textbf{RPDNN-CC}: This is the full model excluding context contents.

\noindent$\bullet$ \textbf{RPDNN-CM}: This is the full model excluding context metadata. 

\noindent$\bullet$ \textbf{RPDNN-Att}: This is the full model excluding attention mechanisms. The last hidden state of LSTM output is used for classification with this setting.
  
\noindent$\bullet$ \textbf{RPDNN-SC-CC}: Only context metadata are used.

\noindent$\bullet$ \textbf{RPDNN-SC-CM}: Only context contents are used.

%% file: sections/results.tex
\section{Results and Discussion}\label{results}
\subsection{Classification Performance}

As shown in Table \ref{results:overall-cvresults} and \ref{results:overall-results}, our proposed model yields SoA performance with larger test data comparable to all the baseline models under two different evaluation techniques while our architecture provides a more abstract context representation and does not specially model many aspects of factuality (e.g., stance, word-level context, sentiment, follower/following relationship, etc.). The full model (RPDNN) achieved an average $F_1$ score of 0.817 in 5-fold CV and that of 0.727 in 7-fold LOO-CV. The result of more stricter LOO-CV shows 7\% improvement over the best comparable SoA method. Details of LOO-CV results are presented in appendix \ref{appendix:loocv_details} In brief, we observe that performance varies slightly for different LOO events. The variance of cross-event performance is 0.0033 in $F_{1}$ and 0.0055 in $Acc.$, which could be attributed to structural issues of different LOO event context rather than actual model capabilities.

\begin{table}
\caption{Comparison of overall CV results.}
\label{results:overall-cvresults}
\footnotesize
\centering
\begin{tabular}{p{\dimexpr 0.4\linewidth-1.2\tabcolsep}
				p{\dimexpr 0.15\linewidth-2.5\tabcolsep}
				p{\dimexpr 0.15\linewidth-2.5\tabcolsep} 
				p{\dimexpr 0.15\linewidth-2.5\tabcolsep} 
				p{\dimexpr 0.15\linewidth-2.5\tabcolsep} }
\bf Methods &  \bf P & \bf R & \bf F1 & \bf Acc. \\ \hline 
\textbf{RPDNN} & 0.768 &	\textbf{0.876} & \textbf{0.817} &	0.803 \\
\textbf{RPDNN-cxt}  & \textbf{0.785} &	0.844 &	0.811 &	\textbf{0.804}\\
\textbf{RPDNN-SC}  & 0.730	& 0.839 &	0.780 &	0.762 \\
\textbf{RPDNN-CC} & 0.762 &	0.846 &	0.801 &	0.788 \\
\textbf{RPDNN-CM} & 0.754 &	0.868	& 0.805 &	0.789 \\
\textbf{RPDNN-Att} & 0.766 &	0.847 &	0.803 &	0.792 \\
\textbf{RPDNN-SC-CM} & 0.779 &	0.733 &	0.754 &	0.762 \\
\textbf{RPDNN-SC-CC} & 0.624 & 0.597 & 0.609 &	0.617\\
\cline{1-5}
\textbf{\cite{zhou2019early}} & \textbf{0.843}\emph{*}  & 0.735\emph{*} & 0.785\emph{*} & \textbf{0.858}\emph{*} \\ 
\textbf{\cite{liu2018early}} & -- & -- & \textbf{0.843} & 0.853 \\
\textbf{\cite{ma2018detect}} & -- & -- & 0.753 & 0.730 \\
\hline
\multicolumn{5}{l}{\emph{*}evaluation metrics are computed over all classes.}
\end{tabular}
\end{table}

\begin{table}
\caption{Comparison of overall LOO-CV results.}
\label{results:overall-results}
\footnotesize
\centering
\begin{tabular}{p{\dimexpr 0.4\linewidth-1.2\tabcolsep}
				p{\dimexpr 0.15\linewidth-2.5\tabcolsep}
				p{\dimexpr 0.15\linewidth-2.5\tabcolsep} 
				p{\dimexpr 0.15\linewidth-2.5\tabcolsep} 
				p{\dimexpr 0.15\linewidth-2.5\tabcolsep} }
\bf Methods &  \bf P & \bf R & \bf F1 & \bf Acc. \\ \hline 
\textbf{RPDNN} & \textbf{0.648} &	0.834 & \textbf{0.727}	& \textbf{0.684} \\
\textbf{RPDNN-cxt}  & 0.626 &	0.838	& 0.715 &	0.667 \\
\textbf{RPDNN-SC}  & 0.621 &	0.796 &	0.694 & 0.648 \\
\textbf{RPDNN-CC} & 0.631 & 	0.800 &	0.705 & 0.654 \\
\textbf{RPDNN-CM} & 0.625	& \textbf{0.862} &	0.723 &	0.669 \\
\textbf{RPDNN-Att} & 0.643	& 0.814 &	0.717 & 0.679 \\
\textbf{RPDNN-SC-CM} & 0.59	& 0.862	& 0.697 & 0.625 \\
\textbf{RPDNN-SC-CC} & 0.568 & 0.519 &	0.514 & 0.544 \\
\cline{1-5}
\textbf{\citelanguageresource{han2019neural}} &  \textbf{0.716} & \textbf{0.614} & \textbf{0.656} & \textbf{0.685} \\   
\textbf{\cite{zubiaga17exploiting}} & 0.692 & 0.559 & 0.601 & -- \\
\hline
\end{tabular}
\end{table}
\textbf{Ablation study observation} The ablation study of the internal baseline models of shows that \textbf{1) source content: } the content of candidate source tweets can be considered as the most important and influential factor in ERD. This observation is consistent with a large body of previous work that exploits source contents alone to measure the credibility of rumors. The source content only model ( ``RPDNN-cxt'') achieved performance comparable to the full model (only 1\% difference with two metrics). The experiment results show that the adoption of the rumor task-specific ELMo model proves to be effective for short-text content embeddings with limited context by capturing significant contextualized representations of rumor-bearing tweets' content. The ELMo embeddings make the most contribution and improve the overall results, which is further supported by  ``RPDNN-SC-CM'' setting. \textbf{2) conversational context: } the context of source tweets can provide additional and effective information to detect rumors. The context-only model ``RPDNN-SC'' achieved comparable performance to the full model (0.780 $F_{1}$ in CV and 0.694 $F_{1}$ in LOO-CV respectively). It is worth noting that our context content only model (``RPDNN-SC-CM'') also achieved SoA performance based on two metrics (0.754 $F_{1}$ in CV and 0.697 $F_{1}$ in LOO-CV). The results indicate that modeling the evolution of public opinion and self-correction mechanism in tweet context is an important and effective approach to ERD. In addition, the metadata only model (``RPDNN-SC-CC'') achieved reasonable performance (0.609$F_{1}$ in CV and 0.514$F_{1}$ in LOO-CV respectively) and incorporating metadata helps to improve precision of full model by 2.3\% with LOO-CV (as observed in ``RPDNN-CM''). This verifies our assumption that the context metadata of rumor source tweets is useful in capturing relevant characteristics of rumor diffusion in early stages. Our observation from the comparative results suggests that although context metadata is more noisy than context content, it can provide effective complementary evidence in the early stages of rumor diffusion with respect to the identification of weak signals. Further experiments can be conducted to investigate its usefulness in cross-platform (i.e., other social media platforms) and cross-language prediction in terms of exploiting a pre-trained metadata model with transfer learning techniques. By comparing ``RPDNN-CC'' and ``RPDNN-CM'' to the full model, the final unified model improves $F_{1}$ performance by around 1-2\%, which can be attributed to its modeling of higher-order feature interactions of two correlated contexts. \textbf{3) context-aware attention mechanisms:} the benefits of incorporating stacked attention mechanisms into a context model are further justified in our experiments by comparison of performance between the full model and attention excluded model ("RPDNN-Att"). Our context-aware attention mechanism can slightly improve both recall and precision, and overall performance with attention achieves a slight improvement in F-measure under the two evaluation settings by 1.4\% and 1\% respectively. Empirical observation in our data indicates that the stacked attention models can reweigh contexts according to their relevance and significance layer by layer. Due to the recurrent structure, the hidden vector close to the end is more informative than its beginning. Thus, for small context, the performance of the attention-based full model is similar to that of the standard LSTM model (i.e., ``RPDNN-Att''). Few representative context samples from the test set with 3 layers of attention weights can be found in Figure \ref{fig:attention_weights_plots} in Appendix. 

\subsection{Training Loss and Performance} 
Based on the experiments, we set the number of epochs to 10 in order to avoid overfitting. Figure \ref{fig:learning_curve_plots} presents training loss and accuracy curve with 10 epochs over time during the training of ``RPDNN'' models in 7-fold LOO-CV. 

\begin{center}
\begin{minipage}{0.48\linewidth}
\includegraphics[width=\linewidth]{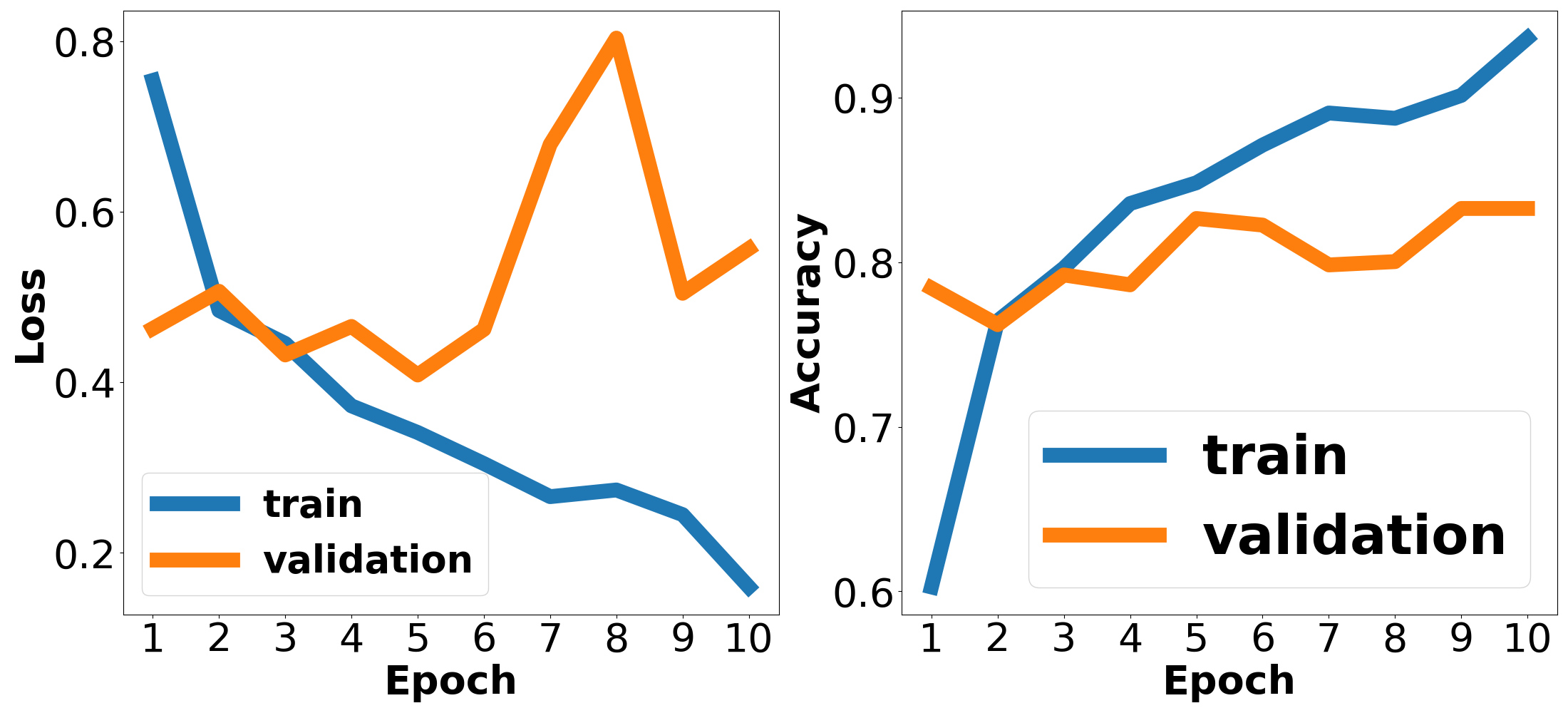}
\captionof{figure}{charliehebdo}
\end{minipage}%
\hfill
\begin{minipage}{0.49\linewidth}
\includegraphics[width=\linewidth]{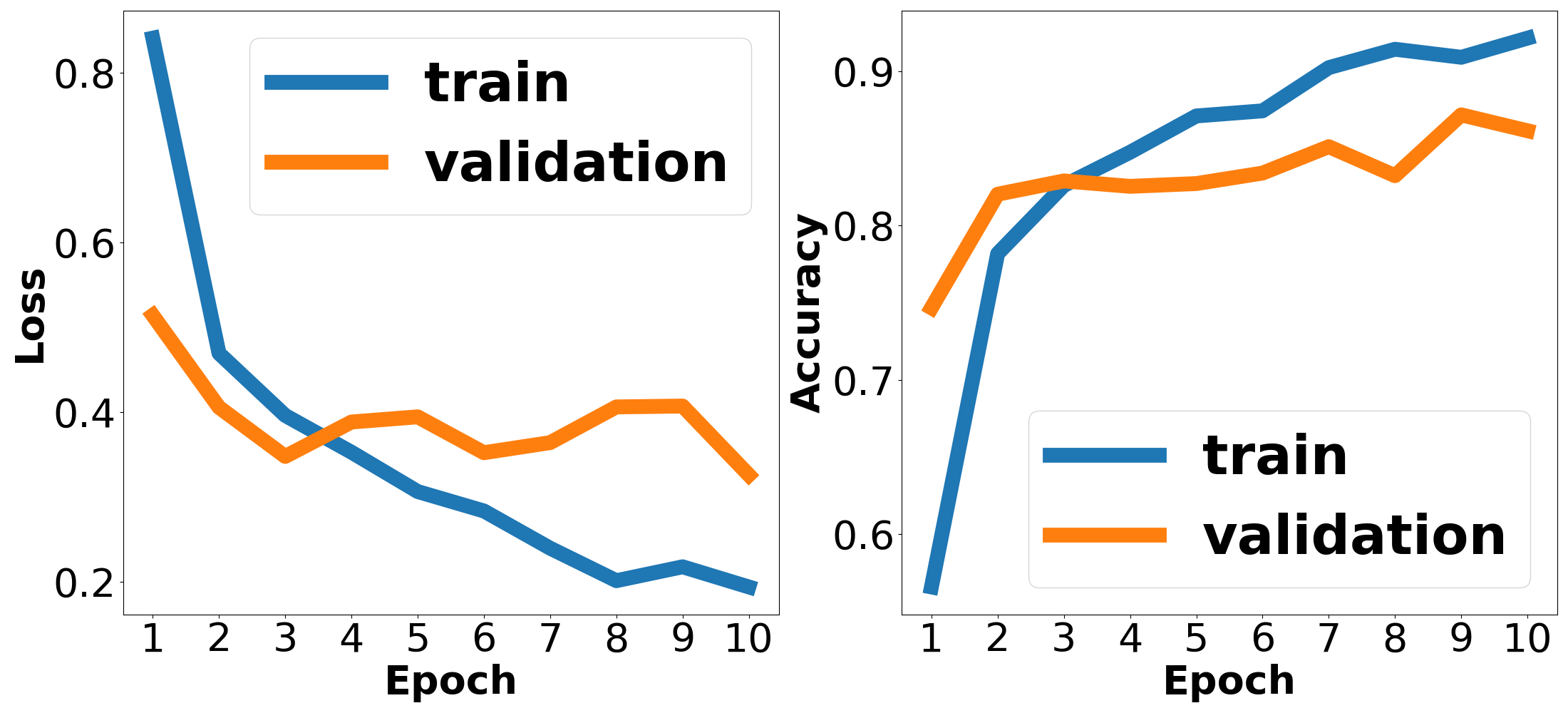}
\captionof{figure}{fergusonunrest}
\end{minipage}
\end{center}
\begin{center}
\begin{minipage}{0.49\linewidth}
\includegraphics[width=\linewidth]{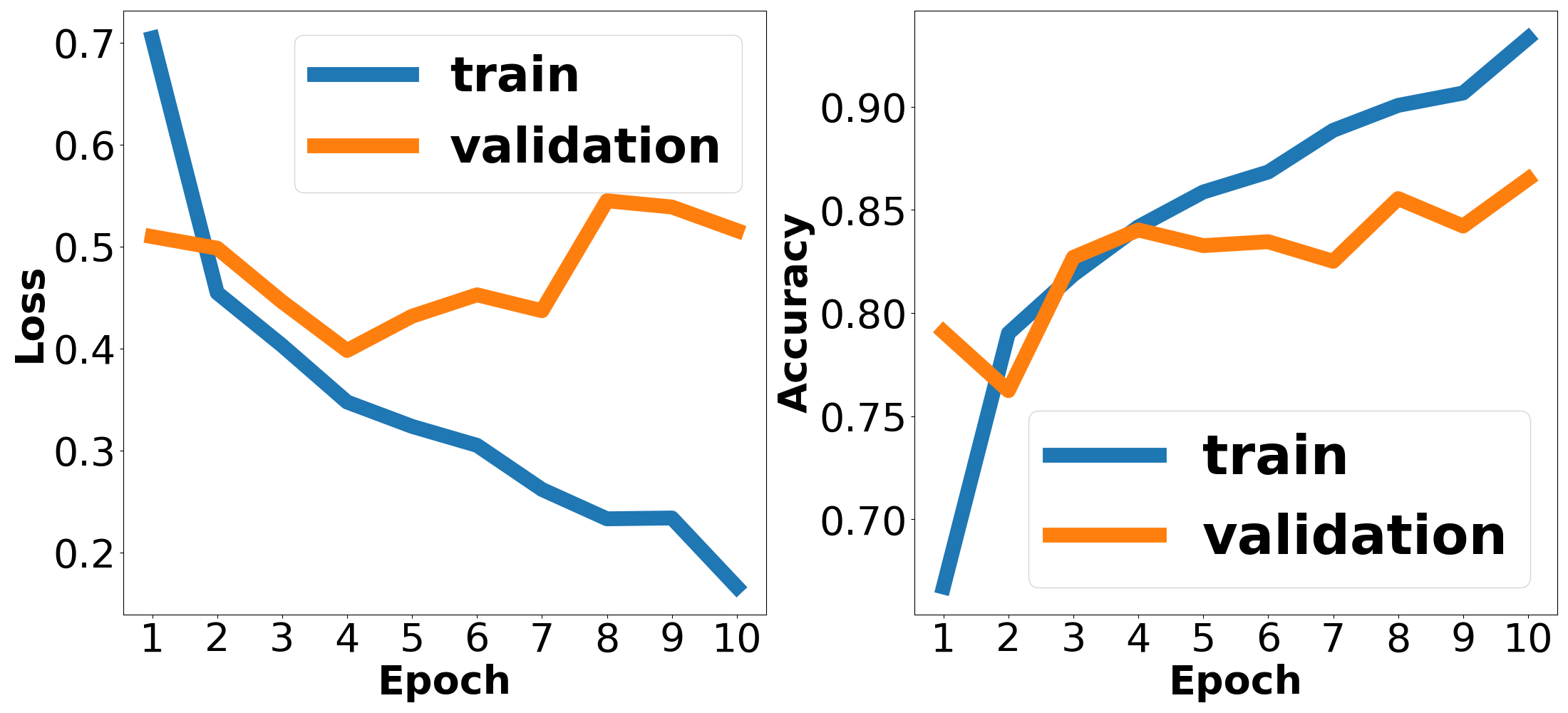}
\captionof{figure}{germanwings}
\end{minipage}
\hfill
\begin{minipage}{0.49\linewidth}
\includegraphics[width=\linewidth]{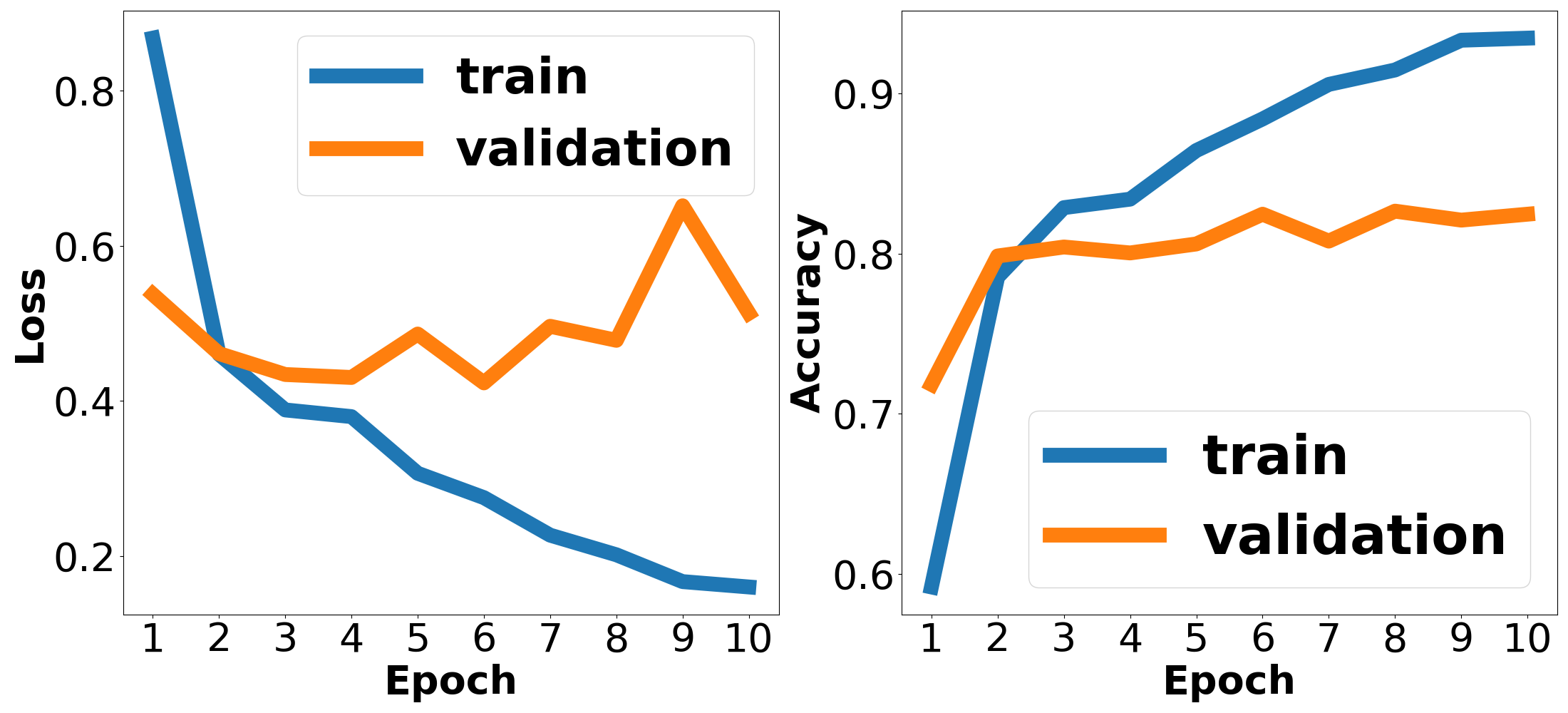}
\captionof{figure}{ottawashooting}
\end{minipage}
\begin{minipage}{0.49\linewidth}
\includegraphics[width=\linewidth]{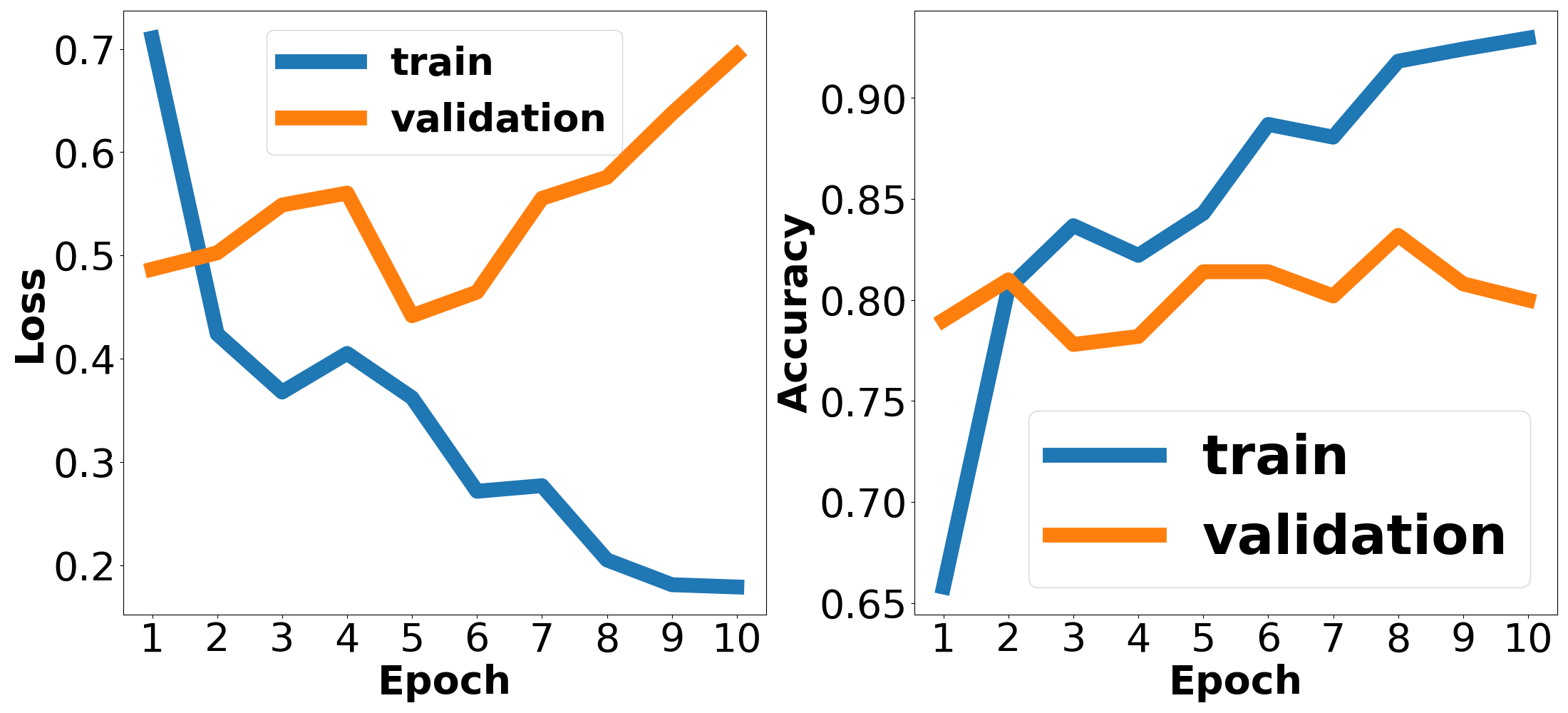}
\captionof{figure}{sydneysiege}
\end{minipage}
\hfill
\begin{minipage}{0.49\linewidth}
\includegraphics[width=\linewidth]{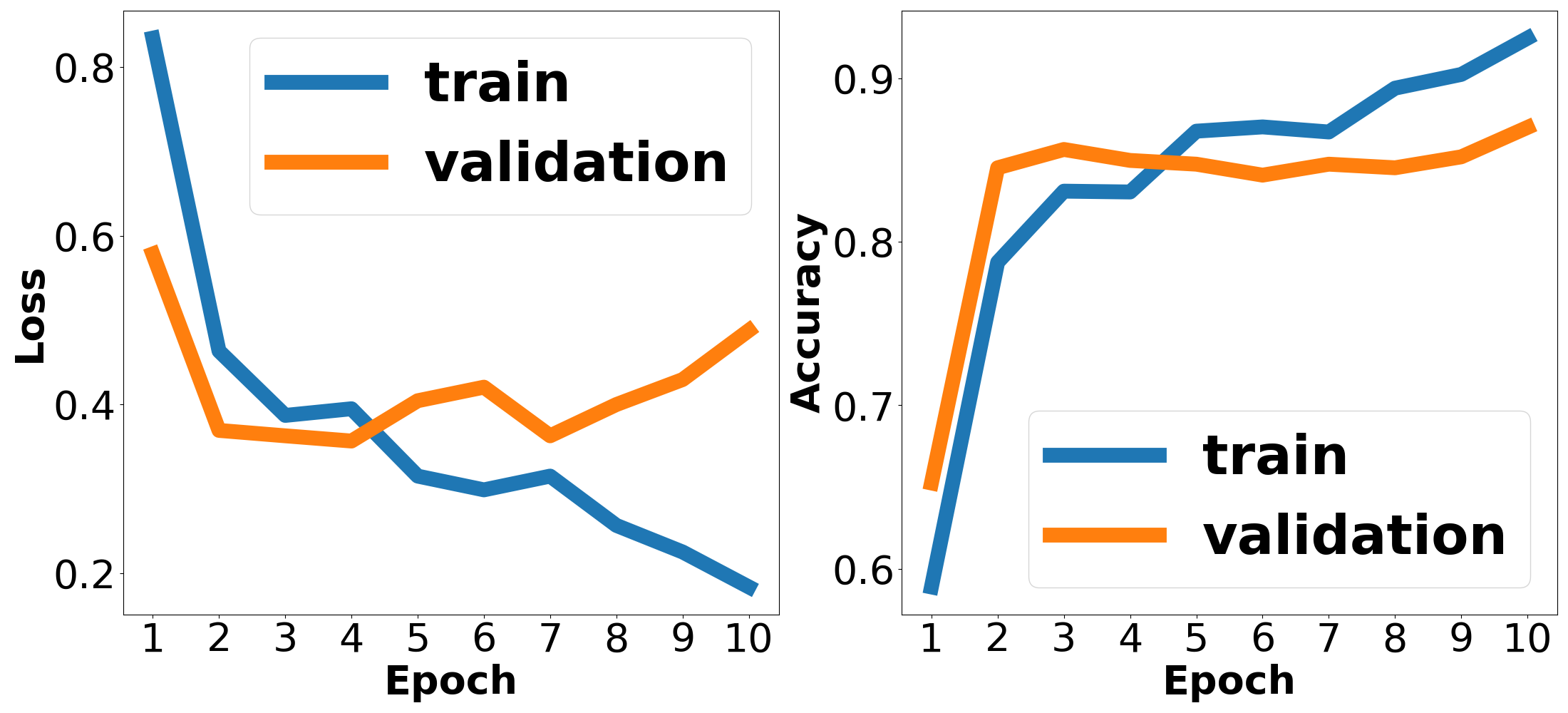}
\captionof{figure}{twitter15}
\end{minipage}
\begin{minipage}{0.49\linewidth}
\includegraphics[width=\linewidth]{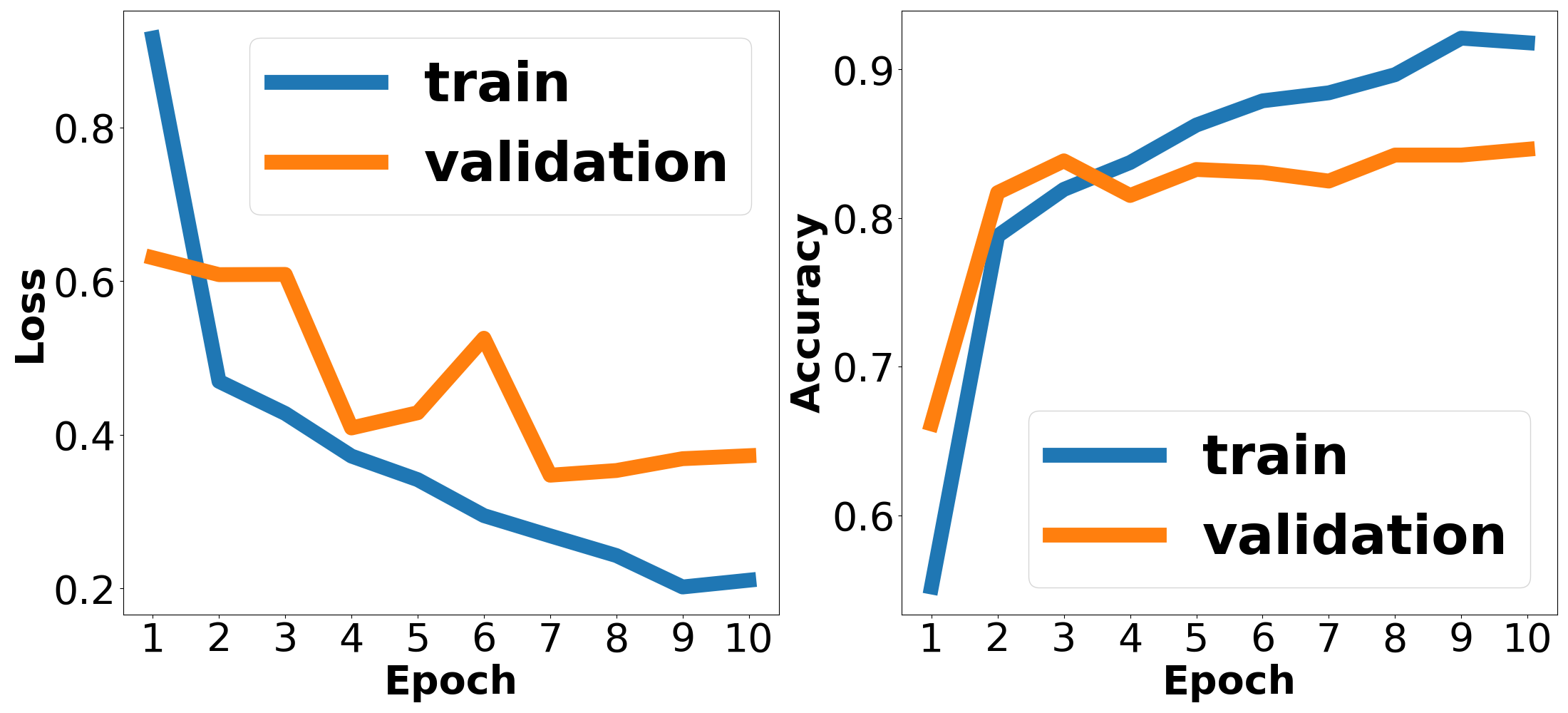}
\captionof{figure}{twitter16}
\end{minipage}
\captionof{figure}{Loss and accuracy curves for 7 folds in LOOCV.}
\label{fig:learning_curve_plots}
\end{center}

The figures on 7 LOO models show steady decreases in training loss within the first 5 epochs and the tendency of overfitting after the 10th epoch. In comparison, we see a constant increase of accuracy in both training and validation sets for all the LOO models. The results show that the ``sydneysiege'' LOO set is the most difficult one to fit. Its divergence in loss can be observed in the very early stage since the 5th epoch and validation accuracy starts to drop after the 10th epoch. The average training time of full models on LOO-CV data is around 28 hours with GPU.

%% file: sections/conclusions.tex
\section{Conclusion}\label{conclusions}
In this paper, we addressed the task of message-level ERD in early development stages of social media rumors where limited information is available. A novel hybrid, context-aware neural network architecture was proposed to learn a unified representation of tweet contents and propagation contexts, enabling the modeling of the evolution of public opinion and the early stages of rumor diffusion. We performed comparative evaluations with two CV techniques and larger test sets from real-life events. The results showed that the proposed model achieves SoA performance. Experimental results showed the advantage of utilizing two types of correlated temporal context inputs from conversational contents and the metadata of tweets in learning an optimal sequential model by improving its effectiveness and generalizability in unseen rumor events. An ablation study proved the positive effect of incorporating a task-specific neural language model and a multi-layered attention model in representation learning in terms of improving resistance to overfitting and noise.

There are several directions for future research. One is to consider the incorporation of social network structure. A potential benefit of modeling retweet chains via follower-following relationship can be studied. In our current work, we find no way to obtain this context data for our public retrospective data using public Twitter API. In addition, the impact of many recent neural language models (typically transformer-based models) and variants of context-aware self-attention models (e.g., multi-head self-attention mechanism in recent work) with larger context size can be examined. Furthermore, generating larger training data with weak supervision technique is promising and can be exploited to allow a deeper NN architecture. It is also interesting to investigate the transferability of a unified model across multiple social media platforms, particularly for the language-independent metadata model. The efficiency and scalability in online social networks are unknown and not examined in this paper.







%% file: sections/appendix_model_setting.tex
\section*{Appendix: Model Settings}\label{appendix_model_settings}
All the parameters of stacked LSTM and attention weights are trained by employing the derivative of the cross-entropy loss function through back-propagation. We use the AdaGrad algorithm for parameter optimisation. As described in Section \ref{method:problemstatement}, source tweets are filtered out based on two constraints: content length and context size. Context sequence size is set to 200 (i.e. $j=200$). The length of each ELMo content embedding is 1024, and that of each metadata feature vector is 27. The number of forward LSTM layers in each stacked LSTM is set to 2, and that of hidden units is set to twice input size. The learning rate and weight decay are set to 1e-4 and 1e-5, respectively. All training instances with corresponding context inputs are iterated over in each epoch where batch size is 128. The number of epochs is set to 10 to avoid overfitting. Leaky ReLU is employed in 3 dense layers. Drop out rates 0.2, 0.3, and 0.3 are respectively applied after each of the three layers. Preliminary results show that the RPDNN suffers from ``dying ReLU'' problem \cite{maas2013rectifier}, which means weights in NNs always drive all inputs to ReLU neurons to negative. This is problematic because ReLu neurons will no longer useful in discriminating the input. Replacing with LReLU fix the problem which gives nonzero gradient for negative value.

%% file: sections/appendix_loocv_details.tex
\section*{Appendix: LOOCV results.}\label{appendix:loocv_details}

Details of LOO-CV results are presented in Table \ref{results:loocv}.

\begin{table*}
	\caption{LOOCV results.}
	\label{results:loocv}
	\footnotesize
	\centering
	\scalebox{0.8}{
	\begin{tabular}{p{\dimexpr 0.2 \linewidth-1.3\tabcolsep}
			p{\dimexpr 0.34 \linewidth-1.3\tabcolsep}
			p{\dimexpr 0.12\linewidth-2.5\tabcolsep}
			p{\dimexpr 0.12\linewidth-2.5\tabcolsep} 
			p{\dimexpr 0.12\linewidth-2.5\tabcolsep} 
			p{\dimexpr 0.12\linewidth-2.5\tabcolsep} 
			p{\dimexpr 0.12\linewidth-2.5\tabcolsep}}
		\bf Event & \bf Models & \bf P & \bf R & \bf F1 & \bf Acc. \\ \hline 
		\multirow{5}{*}{\bf charliehebdo } & RPDNN & 0.743	& 0.882 &0.807 & 0.788 \\
		& RPDNN-cxt & 0.654 & 	\textbf{0.956} &	0.777 & 0.725 \\
		& RPDNN-SC & \textbf{0.754} &	0.759	& 0.757	& 0.756 \\
        & RPDNN-CC & 0.712 &	0.924&	0.804 &	0.698 \\
		& RPDNN-CM & 0.735 &	0.944	& \textbf{0.826}	& \textbf{0.802}  \\
		& RPDNN-Att & 0.751 & 0.868 & 0.805	& 0.79  \\
		& RPDNN-SC-CM & 0.697 &	0.868 & 0.773 & 0.746  \\
		& RPDNN-SC-CC & 0.559 & 0.597 & 0.578 & 0.563 \\
		\cline{2-6}
		& \citelanguageresource{han2019neural} & 0.723 &  0.817 & 0.767 & 0.752  \\
		& CRFs \cite{zubiaga17exploiting} & 0.545 & 0.762 & 0.636 & --  \\ \hline
		
		\multirow{5}{*}{\bf ferguson} & RPDNN & 0.59 & 0.884 & 0.708 & 0.635 \\
		& RPDNN-cxt & 0.564 &	0.781 &	0.655 &	0.588 \\
		& RPDNN-SC & \textbf{0.641} &	0.888	& \textbf{0.745}	& \textbf{0.695}\\
		& RPDNN-CC & 0.567 &	0.798	&0.663	& 0.594\\
		& RPDNN-CM & 0.565 &	\textbf{0.957}	& 0.710 & 0.609 \\
		& RPDNN-Att & 0.627 &	0.67 &	0.647 &	0.635  \\
		& RPDNN-SC-CM & 0.527 &	0.996 & 0.69 & 0.552  \\
		& RPDNN-SC-CC & 0.581 & 0.292 & 0.389 & 0.541 \\
		\cline{2-6}
		& \citelanguageresource{han2019neural} & 0.707 & 0.535 & 0.609 & 0.657 \\
		& CRFs \cite{zubiaga17exploiting} & 0.566 & 0.394 & 0.465 & --  \\  \hline
		
		\multirow{5}{*}{\bf germanwings} & RPDNN & 0.594 &	0.745 &	0.661	 & 0.618 \\
		& RPDNN-cxt & 0.577	 & \textbf{0.887} &	0.699	& 0.618 \\
		& RPDNN-SC & 0.482 & 0.745	& 0.585	& 0.472 \\
		& RPDNN-CC & 0.555 & 0.623 &	0.587 &	0.561 \\
		& RPDNN-CM & 0.556 & 0.708	& 0.622 & 0.571 \\
		& RPDNN-Att & 0.602 & 0.755	&0.67	& 0.627 \\
		& RPDNN-SC-CM & 0.511 & 0.849  & 0.638  & 0.519 \\
		& RPDNN-SC-CC & 0.653  & 0.65  & 0.651  &  0.652 \\
		\cline{2-6}
		& \citelanguageresource{han2019neural} & 0.601 & 0.652 & 0.558 & \textbf{0.630} \\
		& CRFs \cite{zubiaga17exploiting} & \textbf{0.743} & 0.668 & \textbf{0.704} & -- \\ \hline
		
		\multirow{5}{*}{\bf ottawashooting} & RPDNN & 0.647 &	\textbf{0.945}	 & 0.768	 & 0.715 \\
		& RPDNN-cxt & 0.686 &	0.924 &	0.788	& 0.751 \\
		& RPDNN-SC &  0.605 &	0.917 &	0.729 &	0.659 \\
		& RPDNN-CC & 0.743 &	0.879	& \textbf{0.805} &	0.787\\
		& RPDNN-CM & 0.650	& \textbf{0.945} &	0.77 &	0.718 \\
		& RPDNN-Att & 0.652 & 0.914 &	0.761 &	0.713 \\
		& RPDNN-SC-CM & 0.615 & 0.886  & 0.726  & 0.666 \\
		& RPDNN-SC-CC & 0.63 & 0.318 & 0.423 & 0.566 \\
		\cline{2-6}
		& \citelanguageresource{han2019neural} & \textbf{0.85} & 0.71 &  0.77 & \textbf{0.80} \\
		& CRFs \cite{zubiaga17exploiting} & 0.841 & 0.585 & 0.690 & -- \\ \hline

		\multirow{5}{*}{\bf sydneysiege} & RPDNN &\textbf{0.784} &	0.809 & \textbf{0.796}&	\textbf{0.793} \\ 
		& RPDNN-cxt & 0.687 & 	0.861 &	0.764 &	0.734 \\
		& RPDNN-SC & 0.675 & 0.823 & 0.741	& 0.713\\
		& RPDNN-CC & 0.673 & 0.871 & 0.759 &	0.724 \\
		& RPDNN-CM & 0.683 & 0.847 & 0.756 &	0.727  \\
		& RPDNN-Att & 0.684 &	\textbf{0.902} &	0.778	&0.743  \\
		& RPDNN-SC-CM & 0.634  &  0.90  & 0.744 & 0.69 \\
		& RPDNN-SC-CC & 0.68  & 0.366 & 0.476 & 0.597  \\
		\cline{2-6}
		& \citelanguageresource{han2019neural} & 0.755 & 0.644 & 0.695 & 0.717 \\ 
		& CRFs \cite{zubiaga17exploiting} & 0.764 & 0.385 & 0.512 & -- \\ \hline
		
		\multirow{5}{*}{\bf Twitter 15} & RPDNN & 0.59 &	0.79 &	0.676 &	0.621 \\ 
		& RPDNN-cxt & 0.563 &	0.734 &	0.637 &	0.582 \\
		& RPDNN-SC & 0.571 &	0.613 &	0.591 &	0.576 \\
	    & RPDNN-CC & 0.581	&0.731 &	0.648	&0.602\\
		& RPDNN-CM & 0.580 &	\textbf{0.839} &	\textbf{0.686} & 0.616 \\
		& RPDNN-Att & \textbf{0.595} &	0.786 &	0.677 &	\textbf{0.625} \\ 
		& RPDNN-SC-CM & 0.565  &  0.69   & 0.621  & 0.579 \\
		& RPDNN-SC-CC & 0.472  & 0.746 & 0.578  & 0.455 \\
		\hline
		
		\multirow{5}{*}{\bf Twitter 16} & RPDNN & 0.588 &	0.785 &	0.673 &	0.618\\ 
		& RPDNN-cxt & \textbf{0.654} &	0.723	 & 0.687	& 0.67 \\
		& RPDNN-SC & 0.622 & 	\textbf{0.827} &	\textbf{0.71} &	\textbf{0.662} \\
		& RPDNN-CC & 0.585 &	0.775	& 0.667 & 0.613 \\
		& RPDNN-CM & 0.608 &	0.7958 & 0.689 & 0.641  \\
		& RPDNN-Att & 0.589 &	0.801 & 0.679	 & 0.62  \\ 
		& RPDNN-SC-CM & 0.583  &  0.843  & 0.69  &  0.62 \\
		& RPDNN-SC-CC & 0.573  & 0.843  & 0.682  & 0.607 \\
		\hline
	\end{tabular}}
\end{table*}

%% file: sections/appendix_attention_analysis.tex
\section*{Appendix: Analysis of attention degrees}
In Figure \ref{fig:attention_weights_plots}, we present weights of first layers of attention (in ``CC'' and ``CM'' columns) and second layer of attention (in ``CC+CM'' column). The context-level attention weights of example threads are highlighted in different colors according to the rank of their weights in different layers.

\begin{center}
\begin{minipage}{1.00\linewidth}
\includegraphics[clip,scale=0.20, trim=0.5cm 9cm 0.5cm 0.1cm,width=\linewidth]{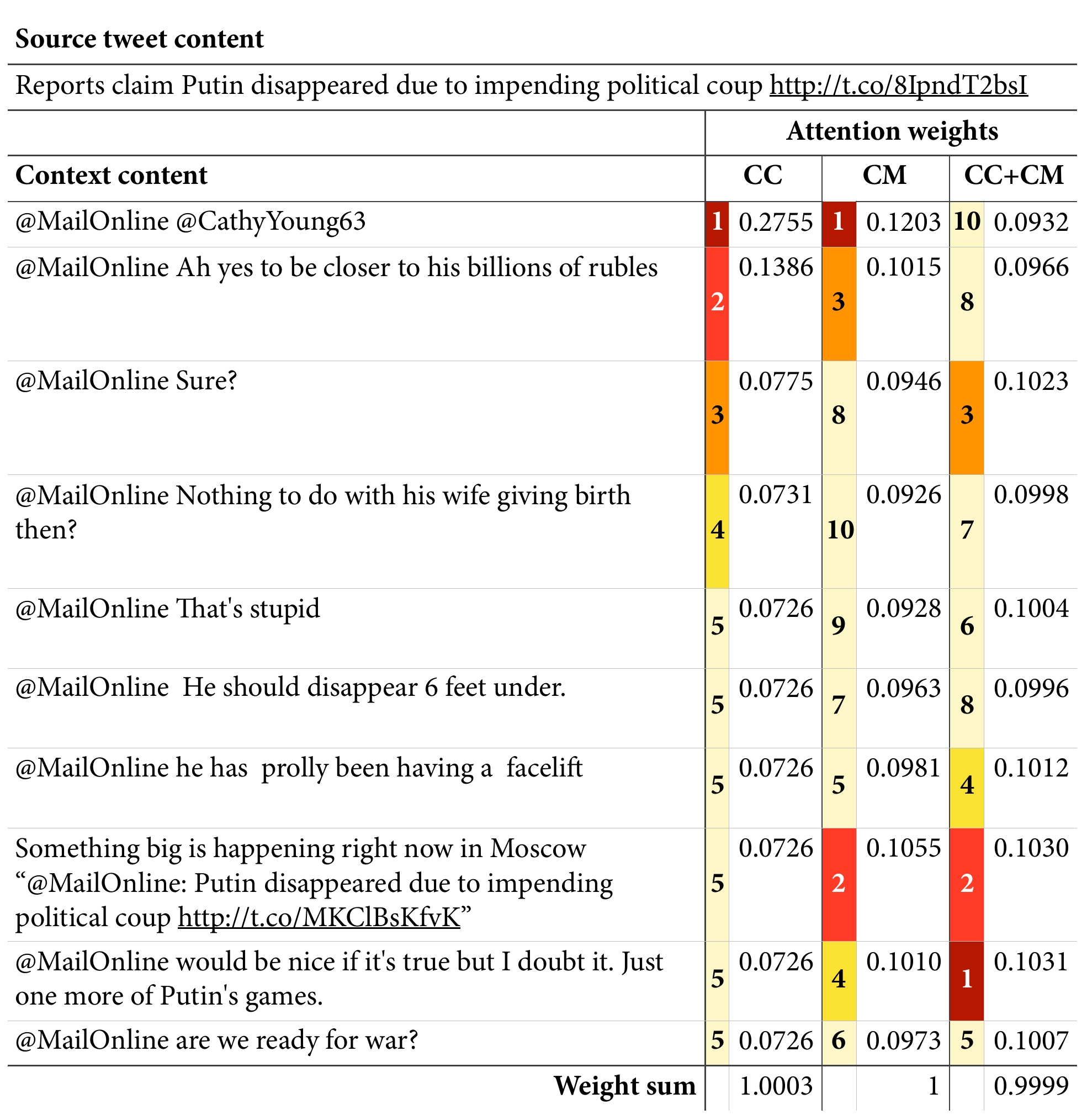}
\includegraphics[clip, trim=0.5cm 12.5cm 0.5cm 0.1cm,width=\linewidth]{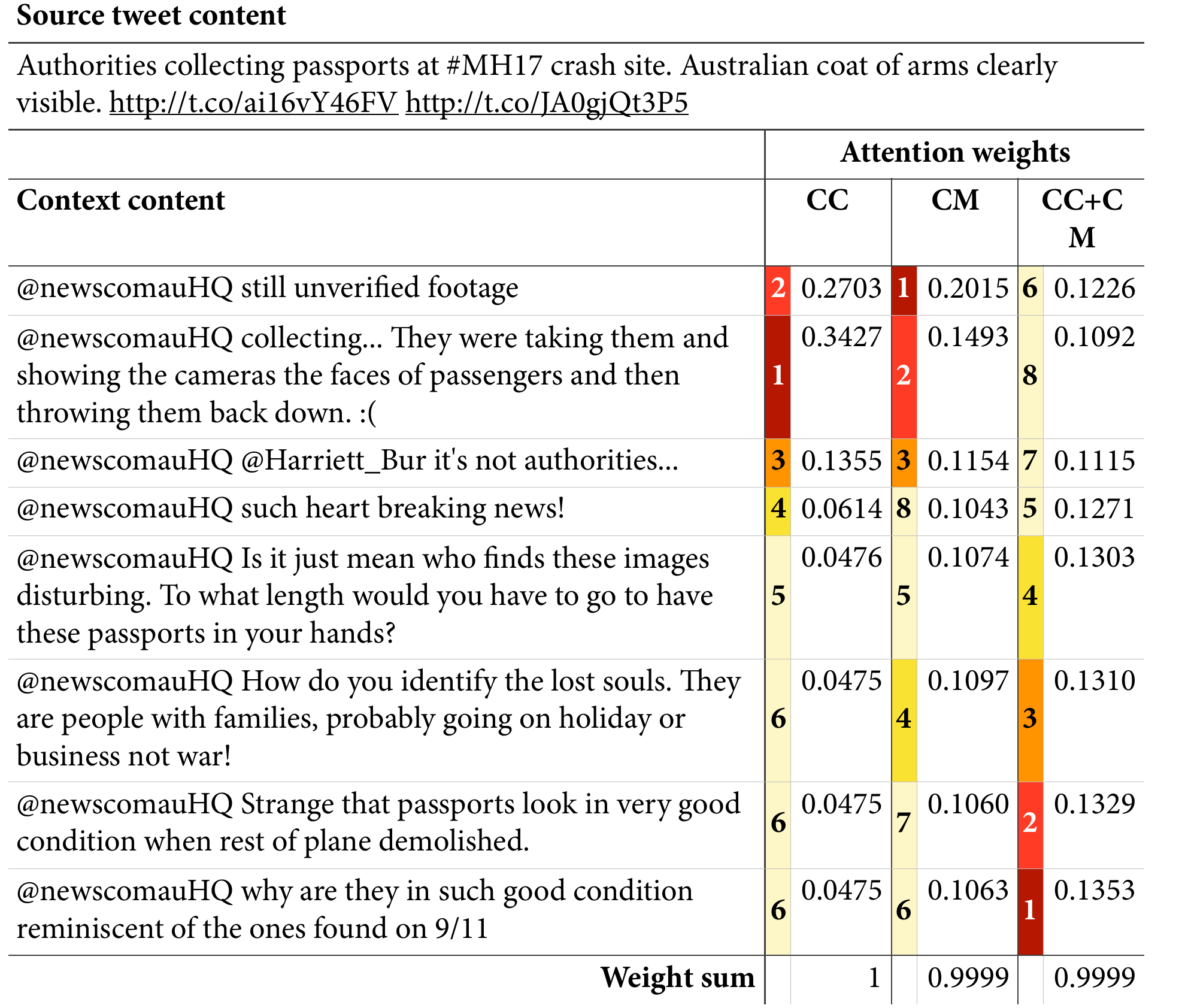}
\end{minipage}%
\captionof{figure}{Visualisation of attention weights for example tweets.}
\label{fig:attention_weights_plots}
\end{center}

The results obtained by the second attention layer (i.e. CC+CM) show that replies expressing doubts and/or questions tend to have higher attention weights. Interestingly, for some replies, the first and second attention layers produce contradictory results, but the latter tends to output more logical results. For instance, the reply ``@MailOnline @CathyYoung63'' in first example of source tweet is in the first rank according to the first layer's results. However, it does not contain any useful information, and its author is not a high-impact user. It is ranked last by the second layer. This observation supports the motivation behind adopting multiple attention layers \cite{yang2016stacked,wang2017residual}, that is, they can progressively refine feature maps and focus on more salient features.
